\documentstyle{article}
\renewcommand{\sin}{\rm{sen}\;}
\begin{document}

\title{O fen\^omeno de quebra de simetria no modelo do oscilador 
anarm\^onico}

\author{A.C. Mastine {\footnote{E-mail: amastine@uel.br}} $\;\;$
e $\;\;$P.L. Natti {\footnote{E-mail: plnatti@uel.br}}\\
Departamento de Matem\'atica, Universidade Estadual de Londrina \\
C. P. 6001, 86051-990, Londrina, PR, Brasil\\
\\
E.R. Takano Natti {\footnote{E-mail: erica.takano@uol.com.br}}\\
Universidade Norte do Paran\'a \\
Campus Londrina, Centro Polit\'ecnico \\
Rua Ti\^ete, 1208 , 86025-230, Londrina, PR, Brasil}


\vskip 1.0cm
\maketitle

\noindent 
{\bf Resumo:} {\it Neste artigo descrevemos atrav\'es de uma t\'ecnica 
n\~ao-perturbativa o problema de condi{\c{c}\~ao} inicial, 
no contexto da Mec\^anica 
Qu\^antica, de um sistema fermi\^onico autointe\-ra\-gente fora 
do equil{\'{\i}}brio na presen{\c{c}a} de um campo magn\'etico. 
Em particular, no regime de campo m\'edio, estudamos o fen\^omeno de 
quebra din\^amica de simetrias neste sistema, identificando os processos 
f{\'{\i}}sicos associados.}

\vskip 0.5cm

\noindent 
{\bf Palavras-chave:} {\it Din\^amica efetiva, quebra de simetria, 
oscilador anarm\^onico fermi\^onico.}

\vskip 0.5cm

\noindent 
{\bf Abstract:} {\it In this article a non-perturbative 
time-dependent technique is used to 
treat the initial value problem, in Quantum Mechanics context, for a 
non-equilibrium self-interacting fermionic system in the presence of 
an external magnetic field. Particularly, in mean-field regime, we 
study the dynamical symmetry breaking phenomenon, identifying the 
physical processes associated.}

\vskip 0.5cm



\section{Introdu{\c{c}\~ao}}

As ci\^encias f{\'{\i}}sicas baseiam-se em modelos matem\'aticos usados 
na descri{\c{c}\~ao} dos fen\^omenos naturais. Na maioria das vezes esta 
descri{\c{c}\~ao} envolve equa{\c{c}\~oes} diferenciais que n\~ao podem ser 
resolvidas analiticamente, isto \'e, n\~ao \'e poss{\'{\i}}vel encontrar 
solu{\c{c}\~oes} destas equa{\c{c}\~oes} em termos de fun{\c{c}\~oes} 
matem\'aticas fundamentais. Normalmente, a descri{\c{c}\~ao} de sistemas 
f{\'{\i}}sicos envolve um n\'umero grande de equa{\c{c}\~oes} acopladas 
que devem ser resolvidas simultaneamente. Nestas situa{\c{c}\~oes} h\'a 
duas poss{\'{\i}}veis 
abordagens, a primeira que consiste em resolver o problema de forma 
num\'erica, atrav\'es de c\'alculos aproximados rea\-lizados em 
computadores; ou resolver o problema estudando algum limite particular 
de interesse f{\'{\i}}sico do sistema, limite em que o sistema tem um 
comportamento simplificado e conseq\"uentemente \'e descrito por 
equa{\c{c}\~oes} diferenciais que podem ser resolvidas analiticamente.

Estamos interessados em procedimentos matem\'aticos que permitam descrever 
o comportamento de sistemas f{\'{\i}}sicos utilizando a segunda abordagem. 
\'E conveniente lembrar que ambas as abordagens s\~ao complementares, 
pois enquanto a abordagem num\'erica permite descrever a evolu{\c{c}\~ao} do 
sistema numa grande variedade de regimes, a abordagem anal{\'{\i}}tica 
simplificada das equa{\c{c}\~oes}, que descreve o sistema num dado regime 
particular, permite interpreta{\c{c}\~oes} que o c\'alculo num\'erico n\~ao 
pode fornecer como veremos neste trabalho.

Abordagens anal{\'{\i}}ticas simplificadas podem ser classificadas em duas 
classes: as perturbativas e as n\~ao-perturbativas. As abordagens 
perturbativas, tamb\'em chamadas apro\-xi\-ma{\c{c}\~oes} perturbativas, 
caracterizam-se por resolverem as equa{\c{c}\~oes} que descrevem a 
evolu{\c{c}\~ao} de sistemas f{\'{\i}}sicos num limite em que um 
par\^ametro f{\'{\i}}sico, que caracteriza o sistema, tende a zero, 
como por exemplo, a intensidade da intera{\c{c}\~ao} entre os 
constituintes do sistema (constante de acoplamento). 
Por outro lado, para sistemas de muitos corpos fortemente acoplados a 
descri{\c{c}\~ao} perturbativa em termos da constante de acoplamento 
n\~ao \'e adequada. Em termos gerais, os m\'etodos 
n\~ao-perturbativos aplicados a tais 
sistemas consistem em buscar uma descri{\c{c}\~ao} qu\^antica completa e 
fechada, mas apenas de uma parte do sistema completo. O intuito desta 
abordagem consiste em simplificar a descri{\c{c}\~ao} do sistema, e com 
isso permitir um tratamento anal{\'{\i}}tico do mesmo. Esta t\'ecnica teve 
origem em problemas onde o sistema de interesse f{\'{\i}}sico estava em 
contato com reservat\'orios t\'ermicos ou ainda em pro\-ble\-mas onde 
t\~ao-somente os elementos diagonais do operador matriz densidade 
completa do sistema eram importantes. Nestes problemas, os graus de 
liberdade do reservat\'orio ou da parte n\~ao-diagonal do operador 
matriz densidade eram considerados irrelevantes e formalmente 
eliminados atrav\'es de um operador de proje{\c{c}\~ao}.

Entretanto, para sistemas de muitos corpos fortemente acoplados \'e 
conveniente ter uma descri{\c{c}\~ao} onde cada parte do sistema \'e 
considerada relevante. Para tais sistemas, Willis e Picard \cite{WP} 
desenvolveram uma t\'ecnica n\~ao-pertubativa 
de proje{\c{c}\~ao} onde os graus de liberdade 
a serem eliminados eram as correla{\c{c}\~oes} (intera{\c{c}\~oes} de 2 
corpos, 3 corpos,..., n corpos), de modo que esta abordagem \'e 
equivalente \`as aproxima{\c{c}\~oes} do tipo campo m\'edio ou do tipo 
Hartree-Fock para sistemas de muitos corpos. O foco deste trabalho 
\'e utilizar esta t\'ecnica para estudar o fen\^omeno de quebra de 
simetria, geralmente associado a processos f{\'{\i}}sicos do tipo 
transi{\c{c}\~oes} de fase, em um sistema de f\'ermions 
autointeragentes na presen{\c{c}a} de um campo magn\'etico externo 
cons\-tante. Na se{\c{c}\~ao} 2 fazemos uma revis\~ao das propriedades 
f{\'{\i}}sicas de um sistema descrito pelo modelo do {\bf{O}}scilador 
{\bf{A}}narm\^onico {\bf{F}}ermi\^onico na presen{\c{c}a} de um campo 
{\bf{M}}agn\'etico (MFAO). Na se{\c{c}\~ao} 3, realizando uma 
transforma\c{c}\~ao do tipo BCS \cite{BCS} e utilizando a t\'ecnica de 
Willis e Picard \cite{WP}, adaptada a sistemas de muitos f\'ermions 
por Toledo Piza e Nemes \cite{Piza}, obt\'em-se as equa\c{c}\~oes 
diferenciais que descrevem a din\^amica efetiva do sistema. 
Na se{\c{c}\~ao} 4 discutimos as simetrias do sistema. Atrav\'es 
de transforma{\c{c}\~oes} particulares do tipo BCS, quebramos 
separadamente estas simetrias, identificando os processos 
f{\'{\i}}sicos associados. 
Finalmente, na se{\c{c}\~ao} 5 reinterpretamos a din\^amica de campo 
m\'edio de nosso sistema e apresentamos as conclus\~oes deste trabalho.

\section{Propriedades f{\'{\i}}sicas do sistema}

M\'etodos aproximativos para tratar problemas de condi{\c{c}\~oes} iniciais 
em teorias qu\^anticas s\~ao essenciais, j\'a que o tratamento exato 
destes problemas raramente \'e poss{\'{\i}}vel, exceto via m\'etodos 
num\'ericos \cite{Erica}. O MFAO \'e um 
modelo exatamente sol\'uvel que pode ser utilizado em v\'arios campos 
da F{\'{\i}}sica, al\'em de ser suficientemente simples, o que faz dele um 
laborat\'orio ideal para o  estudo de novas t\'ecnicas e m\'etodos da 
F{\'{\i}}sica-Matem\'atica.

Come{\c{c}a}mos a se{\c{c}\~ao} descrevendo as principais 
caracter{\'{\i}}sticas do 
modelo. A Hamiltoniana do oscilador a\-nar\-m\^o\-nico fermi\^onico na 
presen{\c{c}a} de um campo magn\'etico externo constante de intensidade $B$ 
\'e dada por 

\begin{equation}
H=\hbar \; \omega \left( a_{1}^{\dag} a_{1} + a_{2}^{\dag} a_{2} \right) 
+U \left( a_{1}^{\dag} a_{1} a_{2}^{\dag} a_{2} \right) + 
g_{_{B}} B \left( a_{1}^{\dag} a_{1} - a_{2}^{\dag} a_{2} \right) 
\;\;,
\end{equation}

\vskip 0.2cm
\noindent
onde $a_{i}^{\dag}$ e $a_{i}$ s\~ao respectivamente operadores 
fermi\^onicos de spin $\;1/2\;$ de cria{\c{c}\~ao} e aniquila{\c{c}\~ao}, 
que satisfazem as rela{\c{c}\~oes} usuais de anticomuta{\c{c}\~ao}, enquanto 
{\'{\i}}ndices $i=1,2$ re\-pre\-sentam res\-pectivamente as poss{\'{\i}}veis 
proje{\c{c}\~oes} de spins $\uparrow$ e $\downarrow$. 
O par\^ametro $U$ representa uma 
intera{\c{c}\~ao} repulsiva entre os el\'etrons do sistema e  $g_{_{B}}$ 
\'e a intensidade de acoplamento do momento magn\'etico intr{\'{\i}}nseco 
(spin) dos f\'ermions com o campo magn\'etico externo $B$. O nome do 
modelo deriva de uma analogia com o modelo do oscilador anarm\^onico 
bos\^onico. 

Com o objetivo de melhor compreender a f{\'{\i}}sica descrita pelo modelo, 
estudemos os seus poss{\'{\i}}veis autoestados e respectivos autovalores 
associados. Sendo $\mid 0\;\rangle$ o v\'acuo do sistema segue que 

\begin{equation}
H \; \mid 0\;\rangle =\left[\hbar \; \omega \left( a_{1}^{\dag} a_{1} + 
a_{2}^{\dag} a_{2} \right) +
U \left( a_{1}^{\dag} a_{1} a_{2}^{\dag} a_{2} \right) + 
g_{_{B}} B \left( a_{1}^{\dag} a_{1} - a_{2}^{\dag} a_{2} \right) 
\right] \mid 0\;\rangle = 
0 \; \mid 0\;\rangle \;\;.
\end{equation}

\vskip 0.2cm
\noindent
Logo, o estado fundamental do sistema (v\'acuo) tem autovalor de energia 
associado nulo. Os demais autoestados da Hamiltoniana s\~ao

\begin{equation}
H \; \left(a_{1}^{\dag} \mid 0\;\rangle \right) = \left[\hbar\omega +
g_{_{B}}\;B \right]\; \left(a_{1}^{\dag} \mid 0\;\rangle \right)
\end{equation}

\vskip 0.2cm

\begin{equation}
H\; \left(a_{2}^{\dag} \mid 0\;\rangle \right)= \left[\hbar\omega - 
g_{_{B}}\;B \right]\; \left(a_{2}^{\dag} \mid 0\;\rangle \right)
\end{equation}

\vskip 0.2cm

\begin{equation}
H\; \left(a_{1}^{\dag}a_{2}^{\dag} \mid 0\;\rangle \right)= 
\left[2\hbar\omega+U\right]\; 
\left(a_{1}^{\dag} a_{2}^{\dag} \mid 0\; \rangle \right)
\;\;.
\end{equation}

\vskip 0.2cm
\noindent
Observe que estes s\~ao todos os poss{\'{\i}}veis autoestados da 
Hamiltoniana (1). Para justificar este fato notamos que 
$H\; \hat O \;a_{i} \mid 0\;\rangle=0 \mid 0\;\rangle$, onde 
$\; \hat O \;$ \'e um operador qualquer, $H \;a_{i}^{\dag}\;
a_{j}a_{j}^{\dag} \; \mid 0\;\rangle=
H \; a_{i}^{\dag} \; \mid 0\;\rangle$ e 
$H \; a_{1}^{\dag}a_{2}^{\dag}\;a_{j}a_{j}^{\dag} \mid 0\;
\rangle =H\; a_{1}^{\dag}a_{2}^{\dag}\; \mid 0\;\rangle$, 
de modo que reca\'{\i}mos nos estados citados acima. Portanto, 
consistentemente com o princ{\'{\i}}pio de Pauli que 
pro\'{\i}be a 
exist\^encia de f\'ermions id\^enticos no mesmo estado 
qu\^antico, nosso sistema tem quatro estados (autoestados) 
que correspondem \`as seguintes configura{\c{c}\~oes}: 

\vskip 0.2cm

i) $\mid 0\;\rangle\;$, estado este que chamamos de v\'acuo, 
o qual n\~ao cont\'em nenhum f\'ermion,

\vskip 0.1cm

ii) $\left(a_{2}^{\dag} \; \mid 0\;\rangle \right)$ estado 
contendo um f\'ermion com spin $\downarrow\;$,

\vskip 0.1cm

iii) $\left(a_{1}^{\dag} \; \mid 0\;\rangle \right)$ estado 
contendo um f\'ermion com spin $\uparrow$ e

\vskip 0.1cm

iv) $\left(a_{1}^{\dag} a_{2}^{\dag} \; \mid 0\;\rangle \right)$ 
estado contendo dois f\'ermion com spins opostos (emparelhados).

\vskip 0.2cm

\noindent
Observe tamb\'em que a presen{\c{c}a} do campo magn\'etico $B$ elimina 
a degeneresc\^encia de energia dos estados $\left(a_{1}^{\dag} 
\mid 0\;\rangle \right)$ e $\left(a_{2}^{\dag} \mid 0\;\rangle 
\right)$, n\~ao afetando os estados $\mid 0\;\rangle$ e 
$\left(a_{1}^{\dag} a_{2}^{\dag} \mid 0\;\rangle \right)$, pois estes 
\'ultimos t\^em momento magn\'etico nulo. Segue que o estado 
inicial mais geral da Hamiltoniana (1), que cont\'em todos 
poss\'{\i}veis autoestados dados em (2-5), \'e dado por 

\begin{equation}
\mid \Psi \; \rangle = \frac{1}{\sqrt N} \left(\rho \; {\hat 1} + 
\beta \; a_{1}^{\dag} 
+ \alpha \; a_{2}^{\dag} + \tau \;a_{1}^{\dag} a_{2}^{\dag} \right)\; 
\mid 0 \;\rangle 
\;\;,
\end{equation}

\vskip 0.2cm
\noindent onde $\rho$, $\beta$, $\alpha$ e $\tau$ s\~ao n\'umeros complexos 
e $N= \rho^{2}+\beta^{2}+\alpha^{2}+\tau^{2}$ \'e a constante de 
normaliza{\c{c}\~ao} do estado. \'E conveniente notar que este estado n\~ao 
tem um n\'umero definido de f\'ermions com spin 
$\uparrow$ e  $\downarrow$ sendo portanto chamado de estado de mistura. 

Como conseq\"u\^encia das considera{\c{c}\~oes} acima, ao contr\'ario do 
Oscilador Anarm\^onico Bos\^onico que apresenta somente solu{\c{c}\~oes} 
apro\-xi\-madas (espectro de energia e estados associados) \cite{Sak}, 
o Oscilador Anarm\^onico Fermi\^onico permite obter solu{\c{c}\~oes} exatas, 
as quais descrevem a intera{\c{c}\~ao}, em um \'unico s{\'{\i}}tio, de dois 
f\'ermios id\^enticos quando sujeitos a um potencial anarm\^onico e a um 
campo magn\'etico externo. 
Finalmente, conv\'em salientar que o modelo MFAO \'e uma vers\~ao 
simplificada dos mo\-de\-los de Hubbard e de Ising \cite{Sal}, 
freq\"uentemente aplicados \`a descri{\c{c}\~ao} de fen\^omenos da 
F{\'{\i}}sica do Estado S\'olido, 
quando reduzidos \`a dimens\~ao espacial zero (um \'unico s{\'{\i}}tio ou 
orbital), de modo que somente os graus de liberdade internos s\~ao 
considerados. Mais detalhes a respeito das aplica{\c{c}\~oes} f{\'{\i}}sicas 
do MFAO podem ser encontradas nas refer\^encias \cite{Sou,MTT}.

Na pr\'oxima se{\c{c}\~ao} atrav\'es de uma transforma{\c{c}\~ao} do tipo 
BCS \cite{BCS}, obteremos a din\^amica efetiva na aproxima{\c{c}\~ao} de 
campo m\'edio para um sistema discreto de f\'ermions descritos pelo MFAO.

\section{Din\^amica efetiva do MFAO}

A Hamiltoniana (1) \'e descrita em termos dos operadores fermi\^onicos 
$a_{i}^{\dag}$ e $a_{i}$ de spin $\;1/2\;$ , que formam uma base 
conhecida como base de part\'{\i}culas. Realizando uma mudan\c{c}a de base  
atrav\'es de uma transforma\c{c}\~ao do tipo BCS, geramos uma din\^amica 
efetiva, a qual minimiza, via um princ\'{\i}pio variacional embutido na 
equa\c{c}\~ao de Heisenberg, a evolu\c{c}\~ao temporal dos observ\'aveis 
de nosso sistema \cite{Mas}. Consideramos a transforma{\c{c}\~ao} definida 
por

\begin{equation}
\lambda_{i}^{\dag}(t)=\sum_{j} \left[\omega_{ji}(t) \; a_{j}^{\dag} 
+ z_{ji}(t) \; a_{j} \right]\;\;,
\end{equation}

\noindent
ou numa forma mais expl{\'{\i}}cita

\begin{equation}
\left[
\begin{array}{c}
\lambda_{1}(t) \\ 
                         \\
\lambda_{2}(t) \\
                            \\
\lambda_{1}^{\dag}(t) \\
                                \\ 
\lambda^{\dag}_{2}(t)
\end{array}
\right]=
\left[
\begin{array}{cccc}
\omega_{11}^{*}(t) &  \omega_{21}^{*}(t)   &     
z_{11}^{*}(t)    &     z_{21}^{*}(t)     \\
                                             
                                   \\
\omega_{12}^{*}(t)  &  \omega_{22}^{*}(t)   &     
z_{12}^{*}(t)    &     z_{22}^{*}(t)      \\
                                                
                                \\
z_{11}(t)    &     z_{21}(t)     &     
\omega_{11}(t)  &  \omega_{21}(t)    \\
                                            
                    \\
z_{12}(t)    &     z_{22}(t)     &     
\omega_{12}(t)  &  \omega_{22}(t)   \\
\end{array}
\right]
\left[
\begin{array}{c}
a_{1}(t)\\
                        \\
a_{2}(t)\\
                        \\
a^{\dag}_{1}(t)\\
                        \\
a^{\dag}_{2}(t)
\end{array}
\right]\;\;, 
\end{equation}
\vskip 0.2cm

\noindent
onde definimos as matrizes $\Omega_{2}$ e $Z_{2}$ como

\begin{equation}
\Omega_{2}=\left[
\begin{array}{cc}
\omega_{11}(t)  &  \omega_{12}(t) \\
                               \\
\omega_{21}(t)  &  \omega_{22}(t)
\end{array}
\right]
\hskip 1.2cm \mbox{e} \hskip 1.2cm 
Z_{2}=\left[
\begin{array}{cc}
z_{11}(t)  &  z_{12}(t)           \\
                               \\
z_{21}(t)  &  z_{22}(t)
\end{array}
\right]\;\;.
\end{equation}
\vskip 0.2cm

\noindent 
Impondo que a transforma{\c{c}\~ao} (8) deva ser unit\'aria, de modo que 
os operadores de campo $\lambda_{i}^{\dag}(t)$ e $\lambda_{i}(t)$, para 
$i=1,2$ , obede\c{c}am as mesmas regras de anticomuta\c{c}\~ao de  
$a_{i}^{\dag}(t)$ e $a_{i}(t)$ a tempos iguais, verificamos que as 
matrizes $\Omega_{2}$ e $Z_{2}$, dadas em (9), necessitam de apenas quatro 
par\^ametros reais independentes para satisfazer as condi{\c{c}\~oes} de 
unitariedade \cite{Mas}. Portanto, em termos dos par\^ametros reais $\theta$, 
$\varphi$, $\gamma$ e $\xi$ construimos $\Omega_{2}$ e $Z_{2}$ como 
\cite{Thomaz}

\begin{eqnarray}
\Omega_{2}&=&\left[
\begin{array}{cc}
\cos{\gamma(t)} \; \cos{\xi(t)}  &  
-\exp{\left[-i\;\varphi(t)\right]}\;\sin{\gamma(t)}\;\cos{\xi(t)} \\
                                             
                                         \\
\exp{\left[i\;\varphi(t)\right]}\;\sin{\gamma(t)}\;\cos{\xi(t)}   &   
\cos{\gamma(t)} \; \cos{\xi(t)} \\
\end{array}
\right] \nonumber
\\  \\ 
Z_{2}&=&\left[
\begin{array}{cc}
\exp{\left[i\;\theta(t)\right]}\;\sin{\gamma(t)}\;\sin{\xi(t)}    
          &  
\exp{\left(i\;[\theta(t)-\varphi(t)]\right)}\;\cos{\gamma(t)}\;
\sin{\xi(t)}   \\
                                                              \\
-\exp{\left(i\;[\theta(t)-\varphi(t)]\right)}\;\cos{\gamma(t)}\;
\sin{\xi(t)}   &
\exp{\left(i\;[\theta(t)-2\varphi(t)]\right)}\;\sin{\gamma(t)}\;
\sin{\xi(t)}
\end{array}
\right]  \nonumber 
\;\;.
\end{eqnarray}
\vskip 0.3cm

\noindent 
Conv\'em salientar que a transforma{\c{c}\~ao} (8) n\~ao \'e \'unica. 
Poder{\'{\i}}amos, por exemplo, ter utilizado fun{\c{c}\~oes} do tipo 
hiperb\'olicas, desde que as condi{\c{c}\~oes} unitariedade fossem 
satisfeitas.

No formalismo de Willis e Picard, as equa\c{c}\~oes de evolu{\c{c}\~ao} 
temporal dos observ\'aveis de um sistema fermi\^onico na aproxima{\c{c}\~ao} 
de campo m\'edio, em termos dos par\^ametros acima definidos, s\~ao 
\cite{Piza}

\begin{eqnarray}
\dot P_{2}+[P_{2},\dot \Omega^{\dag}_{2}\Omega_{2}+\dot Z^{\dag}_{2}Z_{2}]_{-}
&=& -i \; Tr \left([\lambda^{\dag}_{i}\lambda_{j},H]{\cal F}_{0}\right) 
\nonumber\\ \\
\{P_{2},\dot \Omega^{\dag}_{2} Z^{*}_{2}+\dot Z^{\dag}_{2}\Omega^{*}_{2}\}_{+} 
-(\dot \Omega^{\dag}_{2} Z^{*}_{2}+\dot Z^{\dag}_{2}\Omega^{*}_{2}) 
&=& -i \; Tr \left([\lambda_{i}\lambda_{j},H]{\cal F}_{0}\right) \;\;, \nonumber
\end{eqnarray}
\vskip 0.3cm

\noindent
onde a Hamiltoniana $H$ do MFAO deve ser escrita na base de 
quase-part{\'{\i}}culas, utilizando (8) e (10), ${\cal F}_{0}$ \'e a 
matriz densidade de campo m\'edio \cite{Clo} e $P_{2}$ \'e a matriz de ocupa\c{c}\~ao de quase-part\'{\i}culas dada por

\begin{equation}
P_{2}=\langle \lambda^{\dag}_{i}(t)\lambda_{j}(t)\rangle=\left[
\begin{array}{cc}
p_{1}(t)  &            0           \\
                        &                        \\
            0           &  p_{2}(t)
\end{array}
\right]\;\;.
\end{equation}
\vskip 0.3cm

\noindent
Calculando os tra{\c{c}o}s das equa{\c{c}\~oes} de movimento (11), 
obtemos \cite{Mas,Thomaz}

\begin{equation}
\dot p_{1}=0 \hskip 1.5cm \mbox{e} \hskip 1.5cm \dot p_{2}=0
\;\;,
\end{equation}
\vskip 0.2cm

\noindent
indicando que as ocupa{\c{c}\~oes} das quase-part\'{\i}culas (orbitais 
naturais) s\~ao independentes do tempo como esperado na aproxima{\c{c}\~ao} 
de campo m\'edio \cite{Piza}; e 

\begin{eqnarray}
&&-i\;g_{_{B}}\;B\;\left(p_{2}-p_{1}\right)\;
\exp{\left(-i\;\varphi\right)}\;
\sin{(2\;\gamma)}= \nonumber\\ \nonumber\\
&&\left(p_{2}-p_{1}\right)\;
\left[-
\exp{\left(-i\;\varphi\right)}\; \sin{\gamma} \; \cos{\xi} \;\; 
\frac{d}{dt}\left\{\cos{\gamma} \; \cos{\xi}\right\}
\right.
\nonumber \\ \nonumber\\
&& \left. \;\;\;\;\;\;\;\;\;\;\;\;\;\;\;\;
+ \cos{\gamma} \; \cos{\xi} \;\; 
\frac{d}{dt}\left\{\exp{\left(-i\;\varphi\right)} \; \sin{\gamma} \; 
\cos{\xi}\right\}
\right.
\\ \nonumber\\
&& \left. \;\;\;\;\;\;\;\;\;\;\;\;\;\;\;\;
+ \exp{\left[i\;\left(\theta-\varphi\right)\right]}\; \cos{\gamma} 
\; \sin{\xi} \;\; 
\frac{d}{dt}\left\{\exp{\left(-i\;\theta\right)}\;\sin{\gamma} \; 
\sin{\xi}\right\} 
\right.
\nonumber \\ \nonumber\\
&& \left. \;\;\;\;\;\;\;\;\;\;\;\;\;\;\;\;
- \exp{\left[i\;\left(\theta-2\varphi\right)\right]}\; \sin{\gamma} 
\; \sin{\xi} \;\; 
\frac{d}{dt}\left\{\exp{\left[-i\left(\theta-\varphi\right)\right]}
\;\cos{\gamma} 
\; \sin{\xi}\right\} \nonumber
\right] 
\end{eqnarray}

\begin{eqnarray}
&&i\;\frac{1}{2}\;\left(1-p_{1}-p_{2}\right)\;
\left(2\;\hbar\;\omega+U\right)\;
\exp{\left[-i\;\left(\theta-\varphi \right)\right]}\;
\sin{(2\;\xi)} = \nonumber\\ \nonumber\\
&&\left(1-p_{1}-p_{2}\right)\;
\left[
-\exp{\left(i\;\varphi\right)}\; \sin{\gamma} \; \cos{\xi} \;\; 
\frac{d}{dt}\left\{\exp{(-i\;\theta)}\;\sin{\gamma} \; 
\sin{\xi}\right\}
\right.
\nonumber \\ \nonumber\\
&& \left. \;\;\;\;\;\;\;\;\;\;\;\;\;\;\;\;\;\;\;\;\;
- \cos{\gamma} \; \cos{\xi} \;\; 
\frac{d}{dt}\left\{\exp{\left[-i\;\left(\theta-\varphi\right)\right]} \; 
\cos{\gamma} \; \sin{\xi}\right\}
\right.
\\ \nonumber\\
&& \left. \;\;\;\;\;\;\;\;\;\;\;\;\;\;\;\;\;\;\;\;\;
+ \exp{\left[-i\;\left(\theta-\varphi\right)\right]}\; 
\cos{\gamma} \; \sin{\xi} \;\; 
\frac{d}{dt}\left\{\;\cos{\gamma} \; \cos{\xi}\right\} 
\right.
\nonumber \\ \nonumber\\
&& \left. \;\;\;\;\;\;\;\;\;\;\;\;\;\;\;\;\;\;\;\;\;
+ \exp{\left[-i\;\left(\theta-2\varphi\right)\right]}\; \sin{\gamma} 
\; \sin{\xi} \;\; 
\frac{d}{dt}\left\{\exp{\left(-i\;\varphi\right)}\;\sin{\gamma} \; 
\cos{\xi}\right\}
\right] \;\;. \nonumber
\end{eqnarray}
\vskip 0.3cm

\noindent
Resolvendo as derivadas temporais em (14) e (15), e separando as partes 
real e imagin\'aria, obtemos as demais equa{\c{c}\~oes} diferenciais que 
descrevem a din\^amica efetiva de campo m\'edio de nosso sistema, ou seja, 
\vskip -0.3cm
\begin{eqnarray}
&& \dot\gamma=0
\nonumber\\ \nonumber\\
&& \dot\xi=0
\nonumber \\ \\
&& \dot\varphi=2\;g_{_{B}}\;B
\nonumber \\ \nonumber\\
&& \dot\theta=2\;g_{_{B}}\;B+2\; \hbar \; \omega+U
\nonumber \;\;.
\end{eqnarray}
\vskip 0.3cm

Na pr\'oxima se{\c{c}\~ao} iremos interpretar os resultados acima obtidos, 
estudando o fen\^omeno de quebra de simetria neste sistema. 

\section{Quebras de simetrias do sistema}

Para interpretar os resultados obtidos na se{\c{c}\~ao} precedente, 
inicialmente iremos identificar os processos f{\'{\i}}sicos  
presentes na transforma{\c{c}\~ao} BCS que geram as equa{\c{c}\~oes} 
din\^amicas (16). Estudaremos separadamente estes processos de quebra 
de simetrias, interpretando-os. Primeiramente, consideremos a 
transforma{\c{c}\~ao} BCS particular que 
\'e implementada quando impomos que

\begin{equation}
\varphi=0\;\;\;\;\; \mbox{e} \;\;\;\;\; \gamma=0
\end{equation}
\vskip 0.2cm

\noindent
na transforma{\c{c}\~ao} BCS geral (8-10). As matrizes $\Omega_{2}$ e 
$Z_{2}$ escrevem-se agora como

\begin{eqnarray}
\Omega_{2}&=&\left[
\begin{array}{cc}
\cos{\xi}  &  
0                \\
                           \\
0                &   
\cos{\xi}  \\
\end{array}
\right] \nonumber
\\  \\ 
Z_{2}&=&\left[
\begin{array}{cc}
0            &  
\exp{\left(i\;\theta\right)}\;\sin{\xi} \\
                                        \\
-\exp{\left(i\;\theta\right)}\;\sin{\xi} &
0
\end{array}
\right]  \nonumber 
\;\;,
\end{eqnarray}
\vskip 0.3cm

\noindent
de modo que os operadores de cria{\c{c}\~ao} e aniquila{\c{c}\~ao} na 
base de part{\'{\i}}culas e na base de quase-part{\'{\i}}culas 
relacionam-se da seguinte forma

\begin{eqnarray}
\lambda_{1}^{\dag}&=&\left[\cos {\xi} \; a_{1}^{\dag}  + 
\exp{\left(i\;\theta\right)}\;\sin{\xi} \; a_{2} \right] 
\nonumber\\ \nonumber\\ 
\lambda_{2}^{\dag}&=&\left[ 
- \exp{\left(i\;\theta\right)}\;\sin{\xi} \; a_{1} + 
\cos {\xi} \; a_{2}^{\dag} \right]
\nonumber\\ \\ 
\lambda_{1}&=&\left[\cos {\xi} \; a_{1} + 
\exp{\left(-i\;\theta\right)}\;\sin{\xi} \; a_{2}^{\dag}\right]
\nonumber\\ \nonumber\\ 
\lambda_{2}&=&\left[ 
- \exp{\left(-i\;\theta\right)}\;\sin{\xi} \; a_{1}^{\dag}+ 
\cos {\xi} \; a_{2} \right] \nonumber  \;\;.
\end{eqnarray}
\vskip 0.3cm

\noindent

Com o objetivo de identificar as simetrias quebradas devido \`a 
transforma\c{c}\~ao (17-19), verificamos que o operador que conta o 
n\'umero de part{\'{\i}}culas com spin para cima na nova base, 
escreve-se na base de part{\'{\i}}culas como

\begin{equation}
\lambda_{1}^{\dag}\;\lambda_{1} = 
\left(\cos {\xi}\right)^{2}\;a_{1}^{\dag}a_{1} + 
\left(\sin {\xi}\right)^{2}\;a_{2}a_{2}^{\dag} + 
\exp{\left(-i\;\theta\right)}\; \sin {\xi} \; \cos {\xi} \;\;
a_{1}^{\dag}a_{2}^{\dag} +
\exp{\left(i\;\theta\right)} \; \sin {\xi} \; \cos {\xi} \;\;
a_{2}a_{1}\;\;.
\end{equation}
\vskip 0.3cm

\noindent
Notamos que a simetria de n\'umero de part{\'{\i}}culas n\~ao \'e 
conservada, o que pode ser imediatamente observado no terceiro e o 
quarto termos do lado direito da equa\c{c}\~ao (20), onde s\~ao 
respectivamente criadas e aniquiladas duas part{\'{\i}}culas. 
Por outro lado, a simetria de spin permanece intacta, pois todos os 
termos da equa\c{c}\~ao (20) tem spin nulo, consistentemente com o 
fato da parametriza\c{c}\~ao (19) conservar spin. Observando a 
estrutura das matrizes  $\Omega_{2}$ e $Z_{2}$, associamos a 
transforma\c{c}\~ao (17-19) a uma quebra da simetria de rota{\c{c}\~ao} 
em SU$(2)\;$, do tipo emparelhamento, em nosso sistema \cite{Mas,Ryder}.

Calculando as equa{\c{c}\~oes} diferenciais (11) com a 
parametriza{\c{c}\~ao} (18), obtemos a din\^amica efetiva em campo 
m\'edio do sistema devido \`a quebra de simetria de emparelhamento, 
ou seja, 

\begin{eqnarray}
\dot \xi &=& 0
\nonumber\\ \\ 
\dot \theta &=& 2 \; \hbar \; \omega + U
\nonumber  \;\;,
\end{eqnarray}
\vskip 0.3cm

\noindent
onde notamos a presen{\c{c}a} do par\^ametro $U$ da Hamiltoniana (1), 
respons\'avel pela intensidade da autointera{\c{c}\~ao} (emparelhamento) 
entre os f\'ermions. O m\'etodo utilizado acima para a obten{\c{c}\~ao} 
das equa{\c{c}\~oes} din\^amicas (21) \'e tamb\'em chamado de 
aproxima{\c{c}\~ao} do tipo Bogoliubov \cite{BCS}.

Fazemos agora uma segunda parametriza\c{c}\~ao particular, impondo

\begin{equation}
\xi=0\;\;\;\;\; \mbox{e} \;\;\;\;\; \theta=0
\end{equation}
\vskip 0.2cm

\noindent
na transforma{\c{c}\~ao} geral (8-10). Neste caso temos as seguintes 
formas para as matrizes $\Omega_{2}$ e $Z_{2}$

\begin{eqnarray}
\Omega_{2}&=&\left[
\begin{array}{cc}
\cos{\gamma}  &  
-\exp{\left(-i\;\varphi\right)}\;\sin{\gamma} \\
                                                                                      \\
\exp{\left(i\;\varphi\right)}\;\sin{\gamma}  &   
\cos{\gamma}  \\
\end{array}
\right] \nonumber
\\  \\ 
Z_{2}&=&\left[
\begin{array}{cc}
0            &  
0            \\
             \\
0            &
0
\end{array}
\right]  \nonumber 
\;\;,
\end{eqnarray}
\vskip 0.3cm

\noindent
de modo que os operadores de cria{\c{c}\~ao} e aniquila{\c{c}\~ao} 
relacionam-se como

\begin{eqnarray}
\lambda_{1}^{\dag}&=&\left[\cos {\gamma} \; a_{1}^{\dag}  + 
\exp{\left(i\;\varphi\right)}\;\sin{\gamma} \; a_{2}^{\dag} \right] 
\nonumber\\ \nonumber\\ 
\lambda_{2}^{\dag}&=&\left[ 
- \exp{\left(-i\;\varphi\right)}\;\sin{\gamma} \; a_{1}^{\dag} + 
\cos {\gamma} \; a_{2}^{\dag} \right]
\nonumber\\ \\ 
\lambda_{1}&=&\left[\cos {\gamma} \; a_{1} + 
\exp{\left(-i\;\varphi\right)}\;\sin{\gamma} \; a_{2}\right]
\nonumber\\ \nonumber\\ 
\lambda_{2}&=&\left[ 
- \exp{\left(i\;\varphi\right)}\;\sin{\gamma} \; a_{1}+ 
\cos {\gamma} \; a_{2} \right] \nonumber  \;\;.
\end{eqnarray}
\vskip 0.3cm
 
\noindent 
Reescrevendo a partir de (24) o operador que conta o n\'umero de 
part{\'{\i}}culas com spin para cima na nova base, 

\begin{equation}
\lambda_{1}^{\dag}\;\lambda_{1} = 
\left(\cos {\gamma}\right)^{2}\;a_{1}^{\dag}a_{1} + 
\left(\sin {\gamma}\right)^{2}\;a_{2}^{\dag}a_{2} + 
\exp{\left(-i\;\varphi\right)}\; \sin {\gamma} \; \cos {\gamma} \;\;
a_{1}^{\dag}a_{2} +
\exp{\left(i\;\varphi\right)} \; \sin {\gamma} \; \cos {\gamma} \;\;
a_{2}^{\dag}a_{1}\;\;,
\end{equation}
\vskip 0.3cm

\noindent
verificamos imediatamente que o n\'umero de part\'{\i}culas \'e 
conservado, consistentemente com o fato da 
parametriza\c{c}\~ao (24) n\~ao misturar operadores de cria\c{c}\~ao 
com operadores de aniquila\c{c}\~ao. Por outro lado, a simetria de 
spin n\~ao \'e conservada, pois o terceiro e quarto termos da 
equa\c{c}\~ao (25) t\^em spins $+1$ e $-1$, respectivamente. Observando 
$\Omega_{2}$ e $Z_{2}$, associamos a transforma\c{c}\~ao (22-24) a uma quebra 
da simetria de rota{\c{c}\~ao} em SU$(2)\;$, no n\'umero de spin, em nosso 
sistema \cite{Mas,Ryder}. Enfim, calculando neste caso as equa{\c{c}\~oes} 
diferenciais (11), obtemos a din\^amica efetiva em campo m\'edio do sistema 
devido \`a quebra de simetria no n\'umero de spin, ou seja, 

\begin{eqnarray}
\dot \gamma &=& 0
\nonumber\\ \\ 
\dot \varphi &=& 2 \; g_{_{B}} \; B 
\nonumber  \;\;,
\end{eqnarray}
\vskip 0.3cm

\noindent
onde notamos que a autointera{\c{c}\~ao} (emparelhamento) entre os 
f\'ermions do sistema, caracterizada pelo par\^ame\-tro $U$ da 
Hamiltoniana (1), n\~ao est\'a presente nas equa{\c{c}\~oes} din\^amicas 
(26), consistentemente com o fato da simetria de n\'umero de 
part\'{\i}culas permanecer intacta. 
O m\'etodo utilizado acima para a obten{\c{c}\~ao} das 
equa{\c{c}\~oes} din\^amicas (26) \'e tamb\'em chamado de 
aproxima{\c{c}\~ao} do tipo Hartree-Fock \cite{BCS}.

Restam ainda quatro outras poss\'{\i}veis transforma\c{c}\~oes particulares 
do tipo BCS que podem ser implementadas a partir da transforma\c{c}\~ao 
geral (8-10), ou seja, quando $\xi=0\;\; \mbox{e} \;\; \gamma=0 \;$, quando 
$\xi=0\;\; \mbox{e} \;\; \varphi=0 \;$, quando 
$\varphi=0\;\;\ \mbox{e} \;\; \theta=0\;$ e 
quando $\theta=0\;\; \mbox{e} \;\; \gamma=0 \;$. Verifiquemos se 
exis\-tem outros processos f\'{\i}sicos 
(quebra de simetrias) associados a estas 
transforma\c{c}\~oes. Consideramos primeiramente a parametriza\c{c}\~ao 
implementada quando impomos 

\begin{equation}
\xi=0\;\;\;\;\; \mbox{e} \;\;\;\;\; \gamma=0
\end{equation}
\vskip 0.2cm

\noindent
na transforma{\c{c}\~ao} geral (8-10). As matrizes 
$\Omega_{2}$ e $Z_{2}$ adquirem a seguinte estrutura:

\begin{eqnarray}
\Omega_{2}=\left[
\begin{array}{cc}
1  &  0  \\
         \\
0  &  1  \\
\end{array}
\right] 
\;\;\;\;\; \mbox{e} \;\;\;\;\;
Z_{2}=\left[
\begin{array}{cc}
0            &  
0            \\
             \\
0            &
0
\end{array}
\right] 
\;\;,
\end{eqnarray}
\vskip 0.3cm

\noindent
implicando numa transforma\c{c}\~ao identidade. Neste caso, como n\~ao 
ocorre quebra de simetria, n\~ao h\'a uma din\^amica efetiva associada 
(gerada).

Consideremos a seguir a transforma\c{c}\~ao particular BCS implementada 
quando 

\begin{equation}
\xi=0\;\;\;\;\; \mbox{e} \;\;\;\;\; \varphi=0 \;\;.
\end{equation}
\vskip 0.2cm

\noindent
Neste caso as matrizes $\Omega_{2}$ e $Z_{2}$ s\~ao dadas por

\begin{eqnarray}
\Omega_{2}=\left[
\begin{array}{cc}
\cos {\gamma}  &  \sin {\gamma}  \\
         \\
\sin {\gamma}  &  \cos {\gamma}  \\
\end{array}
\right] 
\;\;\;\;\; \mbox{e} \;\;\;\;\;
Z_{2}=\left[
\begin{array}{cc}
0            &      0            \\
             \\
0            &      0
\end{array}
\right] 
\;\;.
\end{eqnarray}
\vskip 0.3cm

\noindent
Observe que a transforma\c{c}\~ao (30) \'e real (ortogonal) e que o 
par\^ametro $\theta$, devido \`a estrutura escolhida para a 
transforma\c{c}\~ao BCS geral (8-10), torna-se arbitr\'ario. Calculando 
as equa{\c{c}\~oes} din\^amicas (11) com esta parametriza\c{c}\~ao, obtemos

\begin{equation}
\dot \gamma = 0 \;\;\;\;\;{\mbox e}\;\;\;\;\;
\sin {2\;\gamma}=0 \;\;,
\end{equation}
\vskip 0.2cm

\noindent
ou ainda,

\begin{equation}
\gamma = k\; \frac{\pi}{2}  \;\;\;\; {\mbox {para}} \;\;\;\;  
k=0,\pm 1,\pm 2,... 
\;\;\;\;\;{\mbox {com}}\;\;\;\;\; 
\theta \;\;{\mbox {arbitr\'ario}}\;\;.
\end{equation}
\vskip 0.2cm

\noindent
Substituindo (32) em (30), notamos que a transforma\c{c}\~ao (29-30), 
dadas as condi\c{c}\~oes iniciais para $\gamma$ e $\theta$, \'e identidade, 
de modo que novamente n\~ao h\'a uma din\^amica efetiva (evolu\c{c}\~ao 
temporal) gerada. Tal situa\c{c}\~ao est\'atica, consistente com o fato da 
transforma\c{c}\~ao ser ortogonal, corresponde simplesmente a uma troca 
de r\'otulos de operadores, dada pelas equa\c{c}\~oes (30-32), n\~ao havendo 
quebra de simetria associada.

Analogamente, a transforma\c{c}\~ao BCS implementada quando 

\begin{equation}
\varphi=0\;\;\;\;\; \mbox{e} \;\;\;\;\; \theta=0 \;\;,
\end{equation}
\vskip 0.2cm

\noindent
gera, a partir de (11), equa\c{c}\~oes est\'aticas para os 
par\^ametros $\gamma$ e $\xi$, ou seja,

\begin{eqnarray}
\dot \gamma &=& 0  \;\;\;\; {\mbox{e}} \;\;\;\;  \sin{2\; \gamma}=0
\nonumber\\ \\ 
\dot \xi &=& 0   \;\;\;\; {\mbox{e}} \;\;\;\;  \sin{2\; \xi}=0
\nonumber  \;\;,
\end{eqnarray}
\vskip 0.3cm

\noindent
cujas solu\c{c}\~oes s\~ao

\begin{eqnarray}
\gamma &=& k\; \frac{\pi}{2}  \;\;\;\; {\mbox{com}} \;\;\;\;  
k=0,\pm 1,\pm 2,...
\nonumber\\ \\ 
\xi &=&  k\; \frac{\pi}{2}  \;\;\;\; {\mbox{com}} \;\;\;\;  
k=0,\pm 1,\pm 2,...
\nonumber  \;\;,
\end{eqnarray}
\vskip 0.3cm

\noindent
n\~ao havendo, portanto, fen\^omeno de quebra de simetria associado.

Enfim, resta-nos estudar a transforma\c{c}\~ao gerada quando tomamos 

\begin{equation}
\gamma=0\;\;\;\;\; \mbox{e} \;\;\;\;\; \theta=0 \;\;
\end{equation}
\vskip 0.2cm

\noindent
em (8-10). Neste caso as matrizes $\Omega_{2}$ e $Z_{2}$ adquirem a seguinte 
estrutura:

\begin{eqnarray}
\Omega_{2}=\left[
\begin{array}{cc}
\cos {\xi}  &  0  \\
         \\
0  &  \cos {\xi}  \\
\end{array}
\right] 
\;\;\;\;\; \mbox{e} \;\;\;\;\;
Z_{2}=\left[
\begin{array}{cc}
0            &      
\exp{\left(-i \varphi \right)}\;\sin{\xi}             \\
             \\
-\exp{\left(-i \varphi\right)}\;\sin{\xi}       &      0
\end{array}
\right] 
\;\;.
\end{eqnarray}
\vskip 0.3cm

\noindent
Calculando as equa{\c{c}\~oes} diferenciais (11) com a 
parametriza{\c{c}\~ao} (36-37), obtemos as equa{\c{c}\~oes} din\^amicas

\begin{eqnarray}
\dot \xi &=& 0
\nonumber\\ \\ 
\dot \varphi &=& 2 \; \hbar \; \omega + B
\nonumber  \;\;,
\end{eqnarray}
\vskip 0.3cm

\noindent
as quais s\~ao id\^enticas \`aquelas da primeira parametriza{\c{c}\~ao}, 
exceto por uma troca de r\'otulo, ou seja, neste caso o par\^ametro 
$\varphi$ desempenha o mesmo papel que o par\^ametro $\theta$ desempenhava 
em (21). Tal afirma{\c{c}\~ao} pode ser comprovada observando-se as 
estruturas das matrizes $\Omega_{2}$ e $Z_{2}$ dadas em (18) e em (37). 
Portanto, podemos concluir que esta transforma{\c{c}\~ao} tamb\'em 
est\'a associada \`a quebra de simetria de n\'umero de part{\'{\i}}culas.

Assim, verificamos que o fen\^omeno de quebra de simetrias de n\'umero de 
part{\'{\i}}culas (f\'ermions) e de n\'umero de spins, implementadas 
atrav\'es da transforma{\c{c}\~ao} do tipo BCS (8-10), s\~ao 
respons\'aveis pela din\^amica efetiva (16) gerada em nosso sistema. Tal 
situa{\c{c}\~ao} tamb\'em \'e observada em teorias qu\^anticas de campos 
\cite{Piza1,Piza2,Plasma1,Plasma2}.

\section{Reinterpreta{\c{c}\~ao} da din\^amica efetiva de campo m\'edio}

Come{\c{c}a}mos rediscutindo a interpreta\c{c}\~ao dada por Thomaz e 
Toledo Piza \cite{Thomaz} para a din\^amica efetiva do MFAO. 
Uma Hamiltoniana efetiva cl\'assica $H_{ef}$, que gera a din\^amica de 
campo m\'edio das vari\'aveis $\;\gamma\;$, $\;\xi\;$, $\;\varphi\;$ e 
$\;\theta\;$ dada em (16), satisfaz, por exemplo, as equa{\c{c}\~oes}

\begin{eqnarray}
&&\dot\varphi=2 \; g_{_{B}} \; B=\;\frac {\partial}
{\partial\gamma}H_{ef}(\;\gamma\;, \;\xi\;, \;\varphi\;, \;\theta\;)
\nonumber\\ \nonumber\\
&&\dot\gamma=0=- \; \frac {\partial}{\partial \varphi}
H_{ef}(\;\gamma\;, \;\xi\;, \;\varphi\;, \;\theta\;) \nonumber\\ \\
&&\dot\theta=2 \; g_{_{B}} \; B+2 \; \hbar \; \omega+U=\;
\frac{\partial}{\partial\xi}
H_{ef}(\;\gamma\;, \;\xi\;, \;\varphi\;, \;\theta\;)\nonumber\\ 
\nonumber\\
&&\dot\xi=0=- \; \frac {\partial}{\partial \theta}
H_{ef}(\;\gamma\;, \;\xi\;, \;\varphi\;, \;\theta\;)\;\;.  \nonumber
\end{eqnarray}
\vskip 0.3cm

\noindent
Integrando as equa{\c{c}\~oes} acima, obtemos de imediato que esta 
Hamiltoniana efetiva cl\'assica \'e dada por 

\begin{equation}
H_{ef}=\left(\;2 \; g_{_{B}} \; B+2 \; \hbar \; \omega+U \;\right) \xi+
\left( \;2 \; g_{_{B}} \; B \;\right) \gamma \;\;.
\end{equation}
\vskip 0.2cm

\noindent
Para interpretar a f{\'{\i}}sica associada \`a Hamiltoniana acima, 
devemos notar que devido \`a escolha da estrutura (39), os 
par\^ametros $\;\varphi\;$ e $\;\theta\;$ correspondem a coordenadas 
generalizadas, enquanto $\;\gamma\;$e $\;\xi\;$ corres\-pondem a momentos 
generalizados. Notando que a Hamiltoniana (40) \'e fun{\c{c}\~ao} 
apenas dos par\^ametros $\;\gamma\;$ e $\;\xi\;$, e que os mesmos 
aparecem acoplados ao campo magn\'etico $B$, conclu\'{\i}mos que 
$\;\gamma\;$ e $\;\xi\;$ podem ser associados a momentos 
magn\'eticos (spins). Para melhor justificar a interpreta{\c{c}\~ao} 
acima, definimos as quantidades 

\begin{eqnarray}
j_{1}(t)=\cos{\gamma(t)} \nonumber\\ \nonumber\\
\alpha_{1}(t)=\theta(t) \nonumber\\ \\
j_{2}(t)=\cos{\xi(t)}    \nonumber\\ \nonumber\\
\alpha_{2}=\varphi(t)  \;\;, \nonumber
\end{eqnarray}
\vskip 0.3cm

\noindent
onde os momentos $j_{1}$ e $j_{2}$ s\~ao conjugados aos \^angulos 
$\theta$ e $\varphi$, respectivamente. Observe que a Hamiltoniana 

\begin{equation}
H_{ef}\left(\alpha_{1},\alpha_{2},j_{1},j_{2} \right)=
\left(\;2 \; g_{_{B}} \; B+2 \; \hbar \; \omega+U \;\right) j_{1} +
\left( \;2 \; g_{_{B}} \; B \;\right) j_{2} 
\end{equation}
\vskip 0.2cm

\noindent
tem as mesmas equa{\c{c}\~oes} de movimento (39). Neste sentido, 
dizemos que as Hamiltonianas dadas em (40) e (42) s\~ao matematicamente 
equivalentes. A partir de (42) verificamos que, dadas as condi{\c{c}\~oes} 
iniciais para $\gamma$ e $\xi\;$, segue que os momentos $j_{1}$ e $j_{2}$ 
s\~ao constantes, pois $\dot\gamma=\dot\xi=0$, assumindo valores tais que

\begin{equation}
-1 \leq j_{1} \leq +1 \;\;\;\;\;\;\; {\mbox{e}} 
\;\;\;\;\;\;\; -1 \leq j_{2} \leq +1 \;\;.
\end{equation}
\vskip 0.2cm

\noindent
Assim, a Hamiltoniana (42), dada em termos de vari\'aveis 
\^angulo-a{\c{c}\~ao},  descreve o movimento angular 
de precess\~ao de duas part{\'{\i}}culas cl\'assicas 
independentes (sem autointera{\c{c}\~ao}) com momentos 
magn\'eticos unit\'arios $\vec j_{1}$ e $\vec j_{2}$ na 
presen{\c{c}a} de um campo magn\'etico externo ${\vec B}$. Nesta 
descri{\c{c}\~ao} as vari\'aveis do tipo a{\c{c}\~ao} $j_{1}$ e 
$j_{2}$, associadas aos \^angulos $\alpha_{1}(t)=\theta(t)$ e
$\alpha_{2}=\varphi(t)$, respectivamente, 
representam as proje{\c{c}\~oes} constantes dos 
momentos angulares de precess\~ao $\vec j_{1}$ e $\vec j_{2}$ 
na dire{\c{c}\~ao} do campo magn\'etico 
externo. Consequentemente, $\theta(t)$ e
$\varphi(t)$ s\~ao os \^angulos longitudinais da 
precess\~ao de $\vec j_{1}$ e $\vec j_{2}$, enquanto as vari\'aveis 
angulares $\gamma$ e $\xi$ s\~ao as colatitudes entre o campo 
magn\'etico externo $\vec B$ constante e os momentos angulares 
de precess\~ao $\vec j_{1}$ e $\vec j_{2}$, respectivamente. 

Portanto, conclu{\'{\i}}mos que a din\^amica efetiva de campo m\'edio de 
um sistema qu\^antico fermi\^onico autointeragente de spin $1/2$ 
descrito pelo MFAO,  
quando sujeito a quebras de simetrias de n\'umero e spin,  
\'e matematicamente id\^entica \`a din\^amica de um sistema cl\'assico 
de dois momentos angulares unit\'arios, independentes, na presen{\c{c}a} 
de um campo magn\'etico externo constante \cite{Thomaz}. 

Em resumo, via um princ{\'{\i}}pio variacional embutido na equa{\c{c}\~ao} 
de Heisenberg, obtivemos a din\^amica efetiva (16) para um sistema 
qu\^antico fermi\^onico de spin $1/2$ descrito pela Hamiltoniana do MFAO 
(1). Tal procedimento foi realizado na aproxima{\c{c}\~ao} de campo m\'edio, 
equivalente \`a aproxima{\c{c}\~ao} de Hartree-Fock-Bogoliubov dependente do 
tempo (TDHFB), onde a varia{\c{c}\~ao} dos par\^ametros da 
transforma{\c{c}\~ao} BCS (8-10) geraram a din\^amica efetiva obtida. 
Verificamos na se{\c{c}\~ao} 4 que os fen\^omenos f{\'{\i}}sicos que 
geraram esta din\^amica efetiva est\~ao associados \`as quebras de 
simetrias de n\'umero de part{\'{\i}}culas (f\'ermions) e de n\'umero de 
spins. Enfim, salientamos que foi devido ao fato de termos realizado um 
tratamento anal{\'{\i}}tico para nosso sistema, que na se{\c{c}\~ao} 5 
foi poss{\'{\i}}vel obter uma equival\^encia matem\'atica entre nosso 
sistema qu\^antico e um sistema cl\'assico. Tais equival\^encias s\~ao 
desej\'aveis, j\'a que permitem uma melhor compreens\~ao de processos 
f{\'{\i}}sicos em sistemas qu\^anticos.

\end{document}